\documentstyle[12pt,a4]{article}

\newcommand\Tr{{\rm tr}}
\newcommand\zed{{\bf Z}}
\newcommand\id{\scriptstyle{\bf 1}}
\begin{document}
\thispagestyle{empty}
\rightline{LMU-HEP-97-28}
\rightline{hep-th/9711039}
\vspace{2.5 truecm}

\centerline{\bf 
STRINGS FROM ORIENTIFOLDS}

\vspace{1.2truecm}
\centerline{\bf Stefan F\" orste and Debashis Ghoshal\footnote{
\noindent
E-mail: \tt Stefan.Foerste@physik.uni-muenchen.de\hfill\\ 
\hphantom{E-mail: }Debashis.Ghoshal@physik.uni-muenchen.de}} 
\vspace{.5truecm}
\centerline{\em Sektion Physik, Universit\"at M\"unchen}
\centerline{\em Theresienstra\ss e 37, 80333 M\"unchen, Germany}

\vspace{2.2truecm}


\vspace{1truecm}

\begin{abstract}
We construct models in 1+1 dimensions with chiral $(0,N)$
supersymmetry by taking orientifolds of type IIB on an eight-torus
identified by different numbers of reflections. The resulting models 
have Dirichlet strings, fivebranes and ninebranes stretched along 
different directions. The cases we study in detail have residual 
chiral supersymmetry $(0,8)$, $(0,4)$ and $(0,2)$. The gravitational 
anomaly in all cases is shown to cancel.
\end{abstract}

\vspace{0.5truecm}

\centerline{PACS: 11.25.M, 04.60.Kz, 11.27.+d}

\vfill

\newpage

\section{Introduction}\label{intro}

Understanding various aspects of compactification of string theory 
is clearly very important. However compactification below four 
dimensions, barring a few notable exceptions, has received little 
attention. A case for three dimensions is that it offers a novel approach 
to the cosmological constant problem\cite{threed}. While the importance 
of two dimensions is due to the fact that here the dimensions of the 
string worldsheet coincide with that of spacetime, and thus one might 
expect to see some new features. Furthermore the large U-duality 
symmetry in low dimensions\cite{ht} means that parameters specifying 
the scale of a given dimension mix with other moduli and the theory, at 
strong coupling, may well have a better description as a higher 
dimensional theory. Indeed strong coupling limit of the type IIA string 
compactified down to two dimensions is related to M-theory in three 
dimensions\cite{stronged} which in turn is related to F-theory  
in four dimensions\cite{vafaf}.

Bagger {\it et al} \cite{bch} considered string compactification on some 
orbifolds of the eight-torus $T^8$. Choosing two of the compactification
radii `big', the resulting string vacua are interpreted as cosmic strings.
The worldsheet of the string stretches along the two non-compact 
dimensions. The specific examples considered were $T^8/\zed_2$, 
$(T^4/\zed_2)\times(T^4/\zed_2)$ and an orbifold limit of a Voisin-Borcea 
type fourfold. 
All these have their holonomy contained in $SU(4)$ and hence type IIB 
(and heterotic) string compactified on these orbifolds give rise to chiral 
supersymmetry in two dimensions:  namely $(0,16)$, $(0,8)$ and $(0,4)$ 
respectively. More recently, motivated by an apparent inconsistency in 
relating to type IIA on an eightfold times a circle, Dasgupta and 
Mukhi\cite{dasmu} studied type IIB compactifications to two 
dimensions with chiral supersymmetry $(0,N)$ with $N=1,2,4,8$ and $16$. 
Topological arguments show that all these lead to consistent string 
backgrounds where the two dimensional gravitational anomaly cancels. 
String compactifications to two dimensions with chiral supersymmetry
were also studied in Ref.\cite{chiral2d}. 

Orbifolds are generalized by orientifolds, where one combines the discrete
identifications in spacetime with a quotient by worldsheet parity 
[\ref{orientfirst}--\ref{orientlast}]. These lead to a more general class of 
string vacua with extended objects, D-branes defined by Dirichlet 
boundary conditions on open strings giving 
additional gauge symmetry. In some cases the orientifolds are known to 
be special points in a more general scheme of compactification of type IIB 
theory, called F-theory, where the dilaton and axion vary in a way 
consistent with the non-perturbative $SL(2,\zed)$ symmetry\cite{sennonp}. 

In this paper, we will construct orientifolds of type IIB string to two 
dimensions by combining worldsheet parity $\Omega$ with spacetime
orbifold $T^8/\zed_2^n$, ($n=1,2,3$) leading to chiral supersymmetry
$(0,2^{4-n})$.  The resulting models are vacua of type
IIB theory with 9-branes, 1-branes and 5-branes, (the latter are not 
present in the $n=1$ case). The 9-branes and 5-branes are 
partly wrapped in the internal space and hence all  branes 
appear as strings in the noncompact
space. In all cases the gravitational anomaly is cancelled
between closed string and open string modes leading to consistent 
string backgrounds. The organization of the paper is as follows: in Section 
\ref{recap} we briefly recall some techniques of orientifolding and some 
special features of two dimensional kinematics; Section \ref{zee2}, 
\ref{zee2sq} and \ref{zee2cub} describe the construction for $n=1,2,3$. In 
Section \ref{zee2} we introduce some notation and explain some common 
features that are assumed in subsequent sections. Section \ref{others} 
outlines the features of the $n=4$ model with $(0,1)$ supersymmetry 
and in Section 
\ref{conclu} we summarize our results and conclude with some outline of
open questions connected to our work.

\section{Orientifold Recap and 2D Kinematics}\label{recap}

An orbifold is a degenerate limit of a manifold that is defined by 
identifying points in space by the elements of a discrete symmetry
group $\Gamma$, an internal symmetry of the worldsheet theory. 
The type IIB theory in addition has a symmetry $\Omega$ that reverses 
the worldsheet parity. The characteristic feature of a new class of string 
vacua called orientifolds is that $\Omega$ is also gauged. This introduces 
the non-orientable Klein bottle diagram (at one loop)  through the 
projection into $\Omega$ invariant states. Nontrivial elements of $\Gamma$ 
give rise to twisted sector states that also propagate in the loop. This 
ensures the modular invariance of the torus amplitude, that at the same 
time eliminates a potential IR divergence.
The Klein bottle diagram is however not protected by any such principle. 
In some cases this divergence can be cancelled by introducing cylinder
and M\"obius strip diagrams (loops of open strings) that also suffer from
similar divergence. Thus the resulting theory is that of unoriented open
and closed strings. The prototype example is the type I theory in ten 
dimensions. 

The open strings may have Neumann or Dirichlet boundary 
conditions. Hence in general open strings have their end points 
restricted to a subspace of spacetime defining extended objects called 
D-branes. The fixed point set of $\Gamma$ defines another kind of defects
called orientifold planes. Far away from the orientifold plane, the theory
appears locally to be one of oriented strings, but close to the fixed plane 
the strings are unoriented. These branes and planes carry charges of the 
gauge fields originating in the Ramond-Ramond sector\cite{polch}. The 
IR divergence can be physically interpreted as due to an exchange of
the corresponding RR fields in the dual closed string tree 
channel\footnote{The divergence actually appears with a prefactor zero
due to residual supersymmetry\cite{polch}, but RR charge conservation 
requires that the RR contribution to the divergence alone cancels.}.
Thus the Klein bottle viewed as a cylinder with crosscap at both ends
corresponds to the interaction between two O-planes, the M\"obius
strip that between an O-plane and a D-brane and the cylinder represents
the interaction between two D-branes. 
The cancellation of the divergence puts constraints on the Chan-Paton
indices that are allowed at the end points of open strings. The 
representations of $\Gamma$ and $\Omega$ on the Chan-Paton labels are
restricted by algebraic consistency conditions\cite{gp}. 

In the type IIA theory, worldsheet parity $\Omega$ is not a symmetry.
However a combination of an odd (even) number of spacetime reflections
together with $\Omega$ is a symmetry of the IIA (respectively IIB) theory. 
This results in D-$p$-branes with $p$ even (odd) appearing in type IIA 
(IIB) theory, and is consistent with their RR field content.

We will construct a class of orientifolds of type IIB theory on $T^8$ with
$\Gamma$ being $\zed_2^n$ for various values of $n$. The details of the
technicalities involved are discussed in \cite{gp} and will not be repeated
here. In brief, the computation is carried out in the loop channel as a
trace:
\begin{equation}
\int {dt\over 2t}\left(\Tr_C \left({1\!+\!\Omega\over 2} {1\! +\! (-)^F\over 2}
(-)^{\scriptstyle{\bf F}}\, \mbox{\bf P} e^{-2\pi t(L_0+{\tilde L}_0)}\right) 
+ \Tr_O\left({1\!+\!\Omega\over 2} {1\! +\! (-)^F\over 2} (-)^{\scriptstyle{\bf F}}\, 
\mbox{\bf P} e^{-2\pi t L_0}\right)\!\right)
\label{loop}
\end{equation}
where $\mbox{\bf P}=\prod(1+R_i)/2$,  ($R_i\in\Gamma$), is the spacetime
part of the orientifold projections, $F$ is the worldsheet fermion number,
and $\mbox{\bf F}$ is the spacetime fermion number. The traces $\Tr_O$ 
and $\Tr_C$ refer to worldsheets with or without boundaries respectively.  
The relevant part of the divergence appear (schematically) as 
follows\cite{caip}:

\medskip
\begin{tabular}{r  l}
Klein bottle: & Closed string NS-NS with $\mbox{\bf P}\Omega (-)^F$\\
M\"obius strip: & Open string R states with $-\mbox{\bf P}\Omega$\\
Cylinder: & Open string NS states with $\mbox{\bf P} (-)^F$
\end{tabular}
\medskip

The resulting models we get live in two (noncompact) dimensional
spacetime. Let $x^{0,1}$ denote these coordinates, $x^\pm=x^0\pm x^1$
are light-cone combinations and $p_{\pm}$ are the corresponding
momenta. In determining the massless spectrum of an orientifold, one
usually imposes the light-cone condition in the noncompact spacetime. 
Two dimensions is special in this respect. Gravity and gauge degrees
of freedom are non-dynamical. There are only bosons and fermions
which can be further divided into purely left-moving (chiral) fields
depending only on $x^-$, and into purely right-moving 
(antichiral) \footnote{In this paper by (anti-)chiral and left/right movers,
we always refer to the two dimensional target space unless stated
otherwise.} ones that depend only on $x^+$. The bosons and fermions
can further be related by bosonization. 

In $D$ dimensions, the light cone condition sets $p_-=0$ (say), 
but for $D=2$ this only gives the right-moving states. The left-movers are
obtained by imposing the other light cone condition $p_+=0$. This is
due to the fact that in two dimensions the left- and right-movers are
truly independent. 

In two dimensions Majorana-Weyl condition can be simultaneously 
imposed on fermions. In general if there are $N_+$ right-moving and 
$N_-$ left-moving supersymmetries, we call the theory $(N_+,N_-)$
supersymmetric. For $N_+\ne N_-$ the theory is chiral and suffers from 
potential gravitational anomaly\cite{agw}. We will mainly be concerned 
with theories with $(0,N)$ supersymmetry. The anomaly cancellation 
condition simplifies to
\begin{equation}
N + {b_- - b_+\over 12} + {f_- - f_+\over 24} = 0.\label{ganomaly}
\end{equation}
Here, $b_\pm$ and $f_\pm$ denote the number of (anti-)chiral bosons
and fermions respectively. In all cases we will discuss the numbers of
chiral and antichiral bosons $b_\pm$ are equal, so only the mismatch
in the fermions will contribute to the anomaly. 

\section{The $T^8/\zed_2$ Orientifold}\label{zee2}

\subsection{Notation and Preliminaries} 

Let us label the coordinates by $x^0,\cdots,x^9$,
with $x^2,\cdots,x^9$ being periodically identified refer to the coordinates
of $T^8$. The discrete group $\Gamma$ here is $\zed_2$ and is 
generated by a reflection ${\cal R}$ of all the compact coordinates. 
To specify the action of ${\cal R}$ on the Ramond sector states, one
represents the reflection in the $ij$ plane as a rotation by $\pm\pi$ 
\begin{equation}
{\cal R} = e^{i\pi\left(-J_{23}+J_{45}+J_{67}-J_{89}\right)},
\end{equation}
where, $J_{ij}$ are the corresponding generators of rotation. The 
worldsheet supercharge defined by 
\begin{equation}
Q = e^{-{\varphi\over 2}}e^{{i\over 2}\left(H_0+H_1+H_2+H_3+H_4\right)}
\label{superch}
\end{equation}
is ${\cal R}$-invariant. The fields $H_k$ are the usual bosonization of the
Ramond sector fermions, $e^{\pm i H_k} = \psi^{2k}\pm i\psi^{2k+1}$; and
$\varphi$ arising from the `bosonization' of the $\beta\gamma$ ghosts. 

The ground state in the Ramond sector is 16-fold degenerate with the
states labelled by 
$e^{-\varphi/2}\exp\left({i\,\mbox{\bf s}\!\cdot\!\mbox{\bf H}}\right)|0\rangle
\equiv |2\mbox{\bf s}\rangle = |2s_0,\cdots,2s_4\rangle$, where $2s_k=\pm$.
The two lightcone choices $p_\pm=0$ for left/right-movers fix the
value of $s_0=\mp{1\over 2}$. Locality with respect to the supercharge
$Q$ in (\ref{superch}), implies that for the `spin' $s_0=+{1\over2}$,
there is an even number of $-$'s in $|2\mbox{\bf s}\rangle$. For  
$s_0=-{1\over2}$, there is an odd number of $+$'s. 

The spacetime reflection ${\cal R}$ leaves $2^8$ points fixed on $T^8$,
giving rise to 256 O(rientifold)-1-planes. This in turn will lead to the 
introduction of open strings with Dirichlet boundary conditions in the
directions $x^2,\cdots,x^9$ defining space-filling D-strings located at
a fixed point in $T^8$. 

Let us remark here that the orbifold $T^8/\zed_2$ cannot be blown up to
a smooth Ricci flat manifold\cite{joyce}. 
Nevertheless (closed, oriented) type IIB string can be 
consistently defined on this orbifold. The resulting model has $(0,16)$ 
supersymmetry, and is conjectured to be dual to M-theory on 
a $T^9/\zed_2$ orbifold\cite{dasmu}.  

\subsection{Massless Spectrum: Closed String Sector}

One gets massless states from closed strings with bosons from the NS-NS
and RR sectors, and fermions from the NS-R sector. Only ${\cal R}$
and $\Omega$ invariant states are allowed. In addition there are open 
string massless bosons (from the NS sector) and fermions (from the
R sector) carrying Chan-Paton labels to be determined by consistency. 
The invariance conditions under ${\cal R}$ and $\Omega$ impose 
restrictions on the Chan-Paton indices.

Let us first determine the right moving modes coming from the untwisted 
closed string sector. The worldsheet left-movers are
\begin{equation}
\begin{array}{r c c c c c}
\mbox{\underline{sector}}&{}&\mbox{\underline{state}}&{}&{}&
\mbox{\underline{${\cal R}$}}\\
\mbox{NS}&{}&\psi^i_{-{1\over 2}}|0\rangle&i=2,\cdots,9&{}&-\\
\mbox{R}&{}&|+,2s_1,\cdots,2s_4\rangle&8s_1s_2s_3s_4=+1&{}&+\\
\end{array}
\label{reigen}
\end{equation}
and similarly for the worldsheet right-movers. Combining the worldsheet
left and right-movers, $\Omega$ invariance keeps 
only symmetric combinations
in the NS-NS sector and antisymmetric ones in the RR sector. This gives
$36+28=64$ antichiral bosons. One does not get any antichiral fermion since
no ${\cal R}$ invariant combination between NS and R sectors is possible.  

The left moving modes are very similar to the above
\begin{equation}
\begin{array}{r c c c c c}
\mbox{\underline{sector}}&{}&\mbox{\underline{state}}&{}&{}&
\mbox{\underline{${\cal R}$}}\\
\mbox{NS}&{}&\psi^i_{-{1\over 2}}|0\rangle&i=2,\cdots,9&{}&-\\
\mbox{R}&{}&|-,2s_1,\cdots,2s_4\rangle&8s_1s_2s_3s_4=-1&{}&-\\
\end{array}
\label{leigen}
\end{equation}
One again obtains $36+28=64$ chiral bosons respectively from the NS-NS
and RR sectors. But now there are also 64 chiral fermions from 
the $\Omega$ invariant NS-R combinations. These fit together into eight 
chiral multiplets\footnote{A chiral multiplet of $(0,N)$ supersymmetry is 
composed of $N$  chiral bosons and $N$ chiral fermions.}
of $(0,8)$ supersymmetry. 

In the twisted sector, the lightest NS state is massive, 
while the only massless 
R state is $|+,0,0,0,0\rangle$. In the end there is no massless mode from the
twisted sector because $\Omega$ invariance removes the only RR state. This
is true of the right-movers, for the left-moving fields due to the GSO 
projection, there is no massless state in the R sector either. 

\subsection{Massless Spectrum: Open String Sector}\label{zed1open}

In the open string NS sector, the massless mode $\psi^i_{-1/2}$ has
$\Omega$ eigenvalue $-1$ ($+1$) for NN (DD) boundary conditions 
respectively. The assignment of the $\Omega$ eigenvalue in the Ramond 
sector follows from the rule that the state $|-----\rangle$ has $\Omega$
eigenvalue $+1$. The occurence of a $+$ in the Ramond state multiplies the
$\Omega$ eigenvalue by $-1$ whenever the corresponding two directions
have NN boundary conditions. With ND (and DN) boundary conditions
the fermions are (half-)integer moded in the NS (respectively R) sector,
however, these are not eigenstates of $\Omega$. 

There are four kinds of open strings --- $99$, $11$, $91$ and $19$ arising 
from different possible boundary conditions. Let us first look at the antichiral
fields. The restriction to be imposed on the 
Chan-Paton factors are as follows
\begin{equation}
\begin{array}{| c | c | c | c |}
\hline
\mbox{B.C.}&\mbox{Sector}&\mbox{Projection}&\mbox{Remark}\\
\hline\hline
\mbox{99}&\mbox{NS}&
\lambda=-\gamma_{\Omega,9}\lambda^T\gamma_{\Omega,9}^{-1}&\mbox{---}\\
{}&{}&=-\gamma_{{\cal R},9}\lambda\gamma_{{\cal R},9}^{-1}&\mbox{---}\\
{}&\mbox{R}&
\lambda=-\gamma_{\Omega,9}\lambda^T\gamma_{\Omega,9}^{-1}&\mbox{---}\\
{}&{}&=+\gamma_{{\cal R},9}\lambda\gamma_{{\cal R},9}^{-1}&\mbox{---}\\
\hline
\mbox{11}&\mbox{NS}&
\lambda=\gamma_{\Omega,1}\lambda^T\gamma_{\Omega,1}^{-1}&\mbox{---}\\
{}&{}&=-\gamma_{{\cal R},1}\lambda\gamma_{{\cal R},1}^{-1}&
\mbox{if located at fixed point}\\
{}&\mbox{R}&
\lambda=-\gamma_{\Omega,1}\lambda^T\gamma_{\Omega,1}^{-1}&\mbox{---}\\
{}&{}&=+\gamma_{{\cal R},1}\lambda\gamma_{{\cal R},1}^{-1}&
\mbox{if located at fixed point}\\
\hline
\mbox{91}&\mbox{R}&
\lambda=+\gamma_{{\cal R},9}\lambda\gamma_{{\cal R},1}^{-1}&
\mbox{if 1-end located at fixed point}\\
\hline
\end{array}
\label{proj2}
\end{equation}
In the above $\gamma_{*,*}$ refer to the representation of the corresponding
projection in the relevant sector\cite{gp}.
Several remarks are in order. Firstly, there is no massless excitation in the
NS sector of the $91$ and $19$ strings, and there is only one (modulo 
Chan-Paton degeneracy) state in the R sector, namely $|+,0,0,0,0\rangle$. 
Secondly, so far we have talked about on-shell degrees of freedom only,
and hence not mentioned the excitations in the $x^\pm$ directions. 
However, to organize states as representations of the `gauge group', it is
useful to note that the Chan-Paton index $\lambda$ of the `vector' 
$\psi^\pm_{-1/2}|0,ij\rangle\lambda_{ji}$ 
(which does not have any dynamics), is subject to the same projections 
as the antichiral fermions. Lastly, since $\Omega$ takes 91 strings to 19 
strings (and vice versa), it does not impose any condition on these strings. 

The condition on the Chan-Paton labels of the chiral fields is very similar,
so instead of repeating the above table, let us point out the differences. 
For the 99 and 11 strings the R sector states are subject to the same 
restriction as the NS sector states. Finally, there is no massless state from 
the 91 and 19 strings.  

\subsection{Tadpole Cancellation}

As we mentioned earlier, three diagrams contribute to the one loop
divergence. The general framework was developped in Ref.\cite{gp} and 
can be applied in our situation in a straightforward way. Hence we will
omit all details of the calculation and only give the divergent contribution 
from each diagram. 

The divergence from the Klein bottle is
\begin{equation}
2^6\left(v_8 + {1\over v_8}\right)\int\limits_{\ell\to\infty}\! d\ell,
\end{equation}
where $v_8$ is the volume of $T^8$ and $t$ in Eq.(\ref{loop}) is related to
$\ell$ by $t=1/4\ell$. The term proportional to $v_8$ comes from the sum
over discrete momenta on $T^8$ and the other term comes from the sum 
over winding modes.
The M\"obius strip contributes
\begin{equation}
- 4\left(v_8\Tr\left(\gamma_{\Omega,9}^{-1}\gamma_{\Omega,9}^T\right)
+ {1\over v_8}\Tr\left(\gamma_{\Omega{\cal R},1}^{-1}
\gamma_{\Omega{\cal R},1}^T\right)\right) \int\limits_{\ell\to\infty}\! d\ell.
\end{equation}
Here, $t=1/8\ell$. 
Finally, the cylinder divergence is
\begin{equation}
{1\over 16}\left[\left(v_8\left(\Tr\gamma_{\id,9}\right)^2 + 
{1\over v_8}\left(\Tr\gamma_{\id,1}\right)^2\right) + 
{1\over 16}\sum_{I=1}^{256}\left(\Tr\gamma_{{\cal R},9} - 
16\Tr\gamma_{{\cal R},I}\right)^2\right]  \int\limits_{\ell\to\infty}\! d\ell,
\end{equation}
where now $t=1/2\ell$, and the sum is over the fixed points of ${\cal R}$.

The divergence can be made to cancel with the following choice
\begin{equation}
\begin{array}{cc}
\gamma_{\Omega,9}=\left(\begin{array}{cc}
                                          1_{16\times 16}&0\\
                                          0&1_{16\times 16}
                                         \end{array}\right)
&\gamma_{{\cal R},9}=\left(\begin{array}{cc}
                                          1_{16\times 16}&0\\
                                          0&-1_{16\times 16}
                                         \end{array}\right)\\
{}&{}\\
\gamma_{\Omega,1}=\left(\begin{array}{cc}
                                          1_{n\times n}&0\\
                                          0&1_{n\times n}
                                         \end{array}\right)
&\gamma_{{\cal R},1}=\left(\begin{array}{cc}
                                          1_{n\times n}&0\\
                                          0&-1_{n\times n}
                                         \end{array}\right)
\end{array}
\label{rep2}
\end{equation}
where, $2n$ is the number of coincident 1-branes of the appropriate type,
and in total there are 32 1-branes (including images). The traces are to
be understood to contain a sum over all types of 1-branes. 

\subsection{Solution of Chan-Paton Matrices and Anomaly Cancellation}

Substituting (\ref{rep2}) in (\ref{proj2}), we find that the Chan-Paton matrices 
of the antichiral fields in the 99 sector are 
\begin{equation}
\mbox{bosons:}\;\lambda=\left(\begin{array}{cc}
                                                 0&B\\
                                                 B^T&0
                                                 \end{array}\right),\;\;\;
\mbox{fermions:}\;\lambda=\left(\begin{array}{cc}
                                                 A_1&0\\
                                                 0&A_2
                                                 \end{array}\right),
\end{equation}
where, $B$ is an arbitrary $16\!\times\! 16$ matrix and $A_i$ 
are antisymmetric
$16\!\times\!  16$ matrices. The fermions therefore transform as the adjoint of
$SO(16)\!\times\!  SO(16)$, which is the `gauge group' and the bosons form a
({\bf 16},{\bf 16}) representation. In the chiral sector both the bosons as well
as the fermions are in the ({\bf 16},{\bf 16}).

For $2a$ D-strings at a fixed point, the solution for the Chan-Paton labels
is the same as in the 99 sector, but with $a\!\times\! a$ blocks. 
Thus the `gauge
group' is $SO(a)\!\times\!  SO(a)$ with the antichiral fermions in the 
adjoint; and the rest in the ({\bf a},{\bf a}).

The open strings stretched between the 9-branes and the $2a$ D-strings at
a fixed point contribute only massless antichiral fermions in the 
({\bf 16},{\bf 1},{\bf a},{\bf 1})+({\bf 1},{\bf 16},{\bf 1},{\bf a}) 
of $SO(16)^2\!\times\! SO(a)^2$. 

If, on the other hand, $2b$ (not counting images) D-strings are off a 
fixed point\footnote{D-strings are allowed to move away from a fixed point in 
packs of two.}, the 
`gauge group' is $SO(2b)$. The antichiral fermions transform in the adjoint, 
and the rest of the fields are in the ({\bf b}({\bf 2b}+{\bf 1})) --- the 
rank-2 symmetric
tensor of $SO(2b)$. Once again the 91 strings contribute only antichiral 
fermions, now in the ({\bf 16},{\bf 1},{\bf 2b})+({\bf 1},{\bf 16},{\bf 2b}) 
of $SO(16)^2\!\times\!  SO(2b)$. 

In a general configuration, there are a bunch (in units of two-packs) of 
D-strings located at a point of $T^8$. Therefore the total number $2a$
of D-strings at fixed points is $a=\sum a_I$, where
$2a_I$ is the number of strings at the $I$th fixed point. Similarly, 
$b=\sum b_J$, where $2b_J$ is the number of strings in
a cluster labelled by $J$. Analogous remarks apply to configurations 
in subsequent sections, but will not be repeated. 

Whenever we move a pair of D-strings away from a fixed point, the 
image also moves. Thus if we have $2a_I$ D-strings at the $I$-th 
fixed point and $2b_J$ coinciding D-strings off the fixed points, we get 
the constraint
\begin{equation}
2\left (\sum_I a_I + 2 \sum_J b_J\right) = 32
\end{equation}
since there are altogether 32 D-strings.

Now we check for the gravitational anomaly of the resulting model. Since
the numbers of chiral and antichiral bosons match, they do not have
any net contribution to the anomaly.  The difference in the numbers of
chiral and antichiral fermions is summarized below
\begin{equation}
\begin{array}{| c || c | c | c | c |}
\hline
\mbox{Sector}&\mbox{closed}&\mbox{99}&\mbox{11}&\mbox{91/19}\\
\hline
f_+ - f_- &-64&-8\cdot 16&-8\cdot\left(a + 2b\right)&
32\cdot\left(a + 2b\right)\\
\hline
\end{array}
\end{equation}
Recall that this model has $(0,8)$ supersymmetry. Now using 
Eq.(\ref{ganomaly}), it is trivial to check that the anomaly vanishes.
This completes our description of the $T^8/\zed_2$ orientifold. 

\section{The $T^8/\zed_2^2$ Orientifold}\label{zee2sq}

\subsection{Preliminaries}

The two $Z_2$ actions are defined by reflections $R_1$ and $R_2$:
\begin{equation}
\begin{array}{c c c c}
R_1:&\left(x^{2,3,4,5},x^{6,7,8,9}\right)&\to&\left(x^{2,3,4,5},-x^{6,7,8,9}\right)\\
R_2:&\left(x^{2,3,4,5},x^{6,7,8,9}\right)&\to&\left(-x^{2,3,4,5},x^{6,7,8,9}\right)
\end{array}
\end{equation}
So the spacetime part describes the orbifold 
$(T^4/\zed_2)\!\times\!  (T^4/\zed_2)$,
which is a degenerate limit of $K3\!\times\!  K3$. The $R_1$ action has $2^4$
fixed points on the second $T^4$, which gives rise to 16 O-$5_1$-planes
wrapped on the first $T^4$. Similarly for $R_2$, 16 O-$5_2$-planes 
localized along $x^{2,3,4,5}$ arise. The combination ${\cal R}=R_1R_2$
yields 256 O-1-planes as in the previous example. 
Thus we need to introduce
Dirichlet 9-, $5_1$-, $5_2$- and 1-branes. Each $Z_2$ action breaks half
of the supersymmetry and $\Omega$ breaks another half, leading to a $(0,4)$
supersymmetric model in two dimensions. 

To define the action of the reflections on the Ramond sector states, we
take the definition
\begin{eqnarray}
R_1 &= e^{i\pi\left(J_{67} - J_{89}\right)}\\
R_2 &= e^{-i\pi\left(J_{23} - J_{45}\right)}
\end{eqnarray}
The supercharge is defined again as in Eq.(\ref{superch}), and is invariant
under $R_1$ and $R_2$. 

\subsection{Massless Spectrum: Closed String Sector}

Let us start with the right-moving modes coming from the untwisted
closed string sector. The left movers on the worldsheet together with
their $R_1$ and $R_2$ eigenvalues are
\begin{equation}
\begin{array}{r c l c c c c}
\mbox{\underline{sector}}&{}&\mbox{\underline{state}}&{}&{}&
\mbox{\underline{$R_1$}}&\mbox{\underline{$R_2$}}\\
\mbox{NS}&{}&\psi^i_{-{1\over 2}}|0\rangle,&i=2,\cdots,5&{}&+&-\\
{}&{}&\psi^i_{-{1\over 2}}|0\rangle,&i=6,\cdots,9&{}&-&+\\
\mbox{R}&{}&|+,2s_1,\cdots,2s_4\rangle,&s_1=s_2,s_3=s_4&{}&+&+\\
{}&{}&|+,2s_1,\cdots,2s_4\rangle,&s_1=-s_2,s_3=-s_4&{}&-&-
\end{array}
\label{reigen2}
\end{equation}
The worldsheet right-movers are similar. The invariant combination from the
NS-NS and RR lead to $20+12=32$ antichiral bosons. No antichiral
fermions survive the projections. 
 
With the light-cone condition $p_+=0$, states and their eigenvalues are
\begin{equation}
\begin{array}{r c l c c c c}
\mbox{\underline{sector}}&{}&\mbox{\underline{state}}&{}&{}&
\mbox{\underline{$R_1$}}&\mbox{\underline{$R_2$}}\\
\mbox{NS}&{}&\psi^i_{-{1\over 2}}|0\rangle,&i=2,\cdots,5&{}&+&-\\
{}&{}&\psi^i_{-{1\over 2}}|0\rangle,&i=6,\cdots,9&{}&-&+\\
\mbox{R}&{}&|-,2s_1,\cdots,2s_4\rangle,&s_1=s_2,s_3=-s_4&{}&-&+\\
{}&{}&|-,2s_1,\cdots,2s_4\rangle,&s_1=-s_2,s_3=s_4&{}&+&-
\end{array}
\label{leigen2}
\end{equation}
Once again, we have 32 chiral bosons, but in addition there are also 32
chiral fermions. These can be combined into eight chiral multiplets of
$(0,4)$ supersymmetry. 

There are massless modes coming from closed strings with twisted boundary
condition under $R_1$ and $R_2$. As before, 
their combined action ${\cal R}$
does not give any massless excitation. Let us focus on the $R_1$ twisted
sector. The necessary ingredients are the right-movers
\begin{equation}
\begin{array}{r c l c c c c}
\mbox{\underline{sector}}&{}&\mbox{\underline{state}}&{}&{}&
\mbox{\underline{$R_1$}}&\mbox{\underline{$R_2$}}\\
\mbox{NS}&{}&|2s_3,2s_4\rangle,&s_3=s_4&{}&+&+\\
\mbox{R}&{}&|+,2s_1,2s_2\rangle,&s_1=-s_2&{}&+&-\\
\end{array}
\label{treigen2}
\end{equation}
and the left-movers
\begin{equation}
\begin{array}{r c l c c c c}
\mbox{\underline{sector}}&{}&\mbox{\underline{state}}&{}&{}&
\mbox{\underline{$R_1$}}&\mbox{\underline{$R_2$}}\\
\mbox{NS}&{}&|2s_3,2s_4\rangle,&s_3=s_4&{}&+&+\\
\mbox{R}&{}&|-,2s_1,2s_2\rangle,&s_1=s_2&{}&+&+\\
\end{array}
\label{tleigen2}
\end{equation}
The invariant combinations are four antichiral bosons and four chiral
bosons and four chiral fermions from each fixed plane of $R_1$. The chiral
fields combine into one chiral multiplet per fixed plane. The $R_2$ twisted
sector gives an exactly identical spectrum. 

Altogether from the closed string sector, one gets 160 antichiral bosons
and 40 chiral multiplets of $(0,4)$ supersymmetry each comprising of four 
chiral bosons and four chiral fermions. 

\subsection{Open String Sector: Massless Spectrum and 
Configuration}\label{zed2open}

In the open string sector massless excitations come from strings stretched
between the various kinds of branes. Let us first concentrate on open
strings stretched between the same kind of branes. The total wavefunction
in the NS sector is $\psi_{-1/2}^i|0,ij\rangle\lambda_{ji}$, and in the
R sector $|2\mbox{\bf s},ij\rangle\lambda_{ji}$. These must be invariant
under $\Omega$ (and $R_1$ and $R_2$ in case the corresponding branes
are at the relevant fixed points). The eigenvalues $\epsilon_\Omega$, 
$\epsilon_{R_1}$ and $\epsilon_{R_2}$ of the relevant modes $\psi$ 
determine the projection conditions on the Chan-Paton matrices
\begin{eqnarray}
\lambda & = &\epsilon_\Omega\gamma_\Omega\lambda^T\gamma_\Omega^{-1}
\nonumber\\
& = &\epsilon_{R_1}\gamma_{R_1}\lambda\gamma_{R_1}^{-1},\quad
\mbox{if the brane is at $R_1$ fixed plane}\\
& = &\epsilon_{R_2}\gamma_{R_2}\lambda\gamma_{R_2}^{-1},\quad
\mbox{if the brane is at $R_2$ fixed plane}\nonumber .
\label{chanpat}
\end{eqnarray}

The eigenvalues of both the left and right-moving NS states are
\begin{equation}
\begin{array}{| c | c c c c c c |}
\hline
\mbox{NS State}&\epsilon_{R_1}&\epsilon_{R_2}&\epsilon_{\Omega,9}&
\epsilon_{\Omega,5_1}&\epsilon_{\Omega,5_2}&\epsilon_{\Omega,1}\\
\hline
\psi^{2,3,4,5}_{-{1\over 2}}&+&-&-&-&+&+\\
\psi^{6,7,8,9}_{-{1\over 2}}&-&+&-&+&-&+\\
\hline
\end{array}
\end{equation}
In the R sector the eigenvalues are different for the left- and right-movers:
\begin{equation}
\begin{array}{| c l | c c c c c c |}
\hline
\mbox{R State}&{}&\epsilon_{R_1}&\epsilon_{R_2}&\epsilon_{\Omega,9}&
\epsilon_{\Omega,5_1}&\epsilon_{\Omega,5_2}&\epsilon_{\Omega,1}\\
\hline
|+,2s_1,\cdots,2s_4\rangle,&s_1=s_2,s_3=s_4&+&+&-&-&-&-\\
|+,2s_1,\cdots,2s_4\rangle,&s_1=-s_2,s_3=-s_4&-&-&-&+&+&-\\
\hline
|-,2s_1,\cdots,2s_4\rangle,&s_1=s_2,s_3=-s_4&-&+&-&+&-&+\\
|-,2s_1,\cdots,2s_4\rangle,&s_1=-s_2,s_3=s_4&+&-&-&-&+&+\\
\hline
\end{array}\label{Rr1r2eigen}
\end{equation}
The assignment of the $\Omega$ eigenvalues has been explained in Section
\ref{zed1open}. The `vector field' which helps 
one in organizing the Chan-Paton
matrices as representations of the `gauge group' has the same eigenvalues
as the first entry in (\ref{Rr1r2eigen}). 

Now we come to open strings between different types of branes. These 
have ND/DN boundary conditions, at least in some directions. With ND/DN
boundary conditions the worldsheet fermions in the NS sector are integer
moded, and half-integer moded in the R sector. For the $95_1$ strings
we have
\begin{equation}
\begin{array}{| r | c l | c c |}
\hline
\mbox{Sector}&\mbox{State}&{}&\epsilon_{R_1}&\epsilon_{R_2}\\
\hline
\mbox{NS}&|2s_3,2s_4\rangle,&s_3=s_4&+&+\\
\hline
\mbox{R}&|+,2s_1,2s_2\rangle,&s_1=-s_2&+&-\\
\cline{2-5}
{}&|-,2s_1,2s_2\rangle,&s_1=s_2&+&+\\
\hline
\end{array}
\end{equation}
The restrictions (\ref{chanpat}) on the Chan-Paton matrices $\lambda$ 
follow from above, but there is no $\Omega$ condition. The $95_2$ strings 
are `morally' similar. Likewise the $5_11$ strings are similar to the $95_2$ 
strings and $5_21$ strings are 
similar to $95_1$ strings. The $R_i$ projection
in (\ref{chanpat}) is imposed if the ends are at the $R_i$ fixed plane. We will
discuss the different possible configurations momentarily. 

Finally, the $91$ and $5_15_2$ strings are similar. Both give rise to the
same massless state: one (modulo Chan-Paton degeneracy) antichiral
fermion. 

Let us now discuss the possible configurations of branes along with the
relevant projections implied on the Chan-Paton labels. The 9-branes
are space-filling, and hence all the projections are imposed on the $99$
strings. The $5_1$-brane is located at a point in the $x^{6,7,8,9}$ directions,
which may or may not be a fixed plane of $R_1$. In either case, $\Omega$ 
and $R_2$ projections are imposed on the $5_15_1$ strings. Now if
($\mbox{A}_1$) the brane is at a fixed plane $R_1$ is imposed in addition;
and if ($\mbox{B}_1$) the brane is off the fixed planes, there is no further
constraint. Analogous 
remarks apply to the $5_2$-branes. For the D-strings, there are four
different locations: (a) at a fixed point of $R_1R_2$; (b) at a fixed plane of
$R_1$, but off $R_2$ fixed planes; (c) at a fixed plane of $R_2$, but off
fixed planes of $R_1$; (d) neither at $R_1$, nor at $R_2$ fixed planes. 
In all cases $\Omega$ constrains the Chan-Paton matrices of the $11$
strings. Further in case (a) $R_1$ and $R_2$, (b) $R_1$, (c) $R_2$ are
imposed. 

For the $95_1$ strings, there is no $\Omega$ projection, and the $R_i$
projections are identical to those of the $5_15_1$ strings depending on 
the location of the $5_1$-end of the string. The $95_2$ strings and $91$
strings are likewise related to the $5_25_2$ and $11$ strings respectively. 

The four possibilities for the $5_15_2$ strings and the corresponding
projections are as follows:
\begin{equation}
\begin{array}{| c || c | c |}
\hline
5_15_2&\mbox{A}_2&\mbox{B}_2\\
\hline
\hline
\mbox{A}_1&R_1, R_2&R_1\\
\hline
\mbox{B}_1&R_2&\mbox{none}\\
\hline
\end{array}
\end{equation}
The different cases of the $5_11$ strings along with the projections to
be imposed are
\begin{equation}
\begin{array}{| c || c | c | c | c |}
\hline
5_11&\mbox{a}&\mbox{b}&\mbox{c}&\mbox{d}\\
\hline
\hline
\mbox{A}_1&R_1, R_2&R_1&\mbox{(massive)}&\mbox{(massive)}\\
\hline
\mbox{B}_1&R_2&\mbox{(massive)}&R_2&\mbox{none}\\
\hline
\end{array}
\end{equation}
In the above we find that in some configurations the lightest
excitations are massive. This is due to the fact that in these 
configurations the branes involved are always separated. 
Finally, the different 
configurations of $5_21$ strings are analogous to the $5_11$ case.

\subsection{Tadpole Cancellation}

The divergence from the Klein bottle diagram comes from four sources
\begin{equation}
2^5\left(v^{(1)}_4v^{(2)}_4 + {v^{(2)}_4\over v^{(1)}_4} +  
{v^{(1)}_4\over v^{(2)}_4} + {1\over v^{(1)}_4v^{(2)}_4} \right)
\int\limits_{\ell\to\infty}\! d\ell,
\end{equation}
where $v_4^{(i)}$ are the volumes of the two $T^4$'s. The contribution
from the M\"obius strip is
\begin{eqnarray}
-2\Bigg( v^{(1)}_4v^{(2)}_4\Tr\left(\gamma^{-1}_{\Omega,9} 
\gamma^T_{\Omega,9}\right)  &+& {v^{(2)}_4\over v^{(1)}_4}
\Tr\left(\gamma^{-1}_{\Omega R_1,5_1}\gamma^T_{\Omega R_1,5_1}\right) 
+ {v^{(1)}_4\over v^{(2)}_4}\Tr\left(\gamma^{-1}_{\Omega R_2,5_2}
\gamma^T_{\Omega R_2,5_2}\right) \nonumber\\
&+& {1\over v^{(1)}_4v^{(2)}_4} 
\Tr\left(\gamma^{-1}_{\Omega R_1R_2,1}\gamma^T_{\Omega R_1R_2,1}\right) 
\Bigg)\int\limits_{\ell\to\infty}\! d\ell,
\end{eqnarray}
And finally the cylinder gives
\begin{eqnarray}
{1\over 2^5} &\times &\left[ v^{(1)}_4v^{(2)}_4\left(\Tr\gamma_{\id,9}\right)^2 +
{v^{(2)}_4\over v^{(1)}_4} \left(\Tr\gamma_{\id,5_1}\right)^2 +
{v^{(1)}_4\over v^{(2)}_4} \left(\Tr\gamma_{\id,5_2}\right)^2 +
{1\over v^{(1)}_4v^{(2)}_4} \left(\Tr\gamma_{\id,1}\right)^2\right. \nonumber\\
&+&{v_4^{(2)}\over 4}\sum_{I=1}^{16}\left(\Tr\gamma_{R_1,9} -
4\Tr\gamma^{(I)}_{R_1,5_1}\right)^2 +
{v_4^{(1)}\over 4}\sum_{I=1}^{16}\left(\Tr\gamma_{R_2,9} -
4\Tr\gamma^{(I)}_{R_2,5_2}\right)^2 \nonumber\\
&+&{1\over 4 v_4^{(2)}}\sum_{I=1}^{16} \left(\Tr\gamma_{R_1,5_2} -
4\Tr\gamma^{(I)}_{R_1,1}\right)^2 + 
{1\over 4 v_4^{(1)}}\sum_{I=1}^{16} \left(\Tr\gamma_{R_2,5_1} -
4\Tr\gamma^{(I)}_{R_2,1}\right)^2 \nonumber\\
&+&\left. {1\over 16}\sum_{I=1}^{256}\left(\Tr\gamma_{R_1R_2,9} - 16
\Tr\gamma_{R_1R_2,1}^{(I)}\right)^2 +
\sum_{I=1}^{16}\sum_{J=1}^{16}\left(\Tr\gamma_{R_1R_2,5_1}^{(I)} -
\Tr\gamma_{R_1R_2,5_2}^{(J)}\right)^2\right],
\end{eqnarray}
times the divergent integral $\int\limits_{\ell\to\infty}\! d\ell$.

The contributions from the Klein bottle, M\"obius strip and the first
line of the cylinder add up to a perfect square for each type of brane.
The total divergence cancels when $\gamma_{\Omega,9}$, 
$\gamma_{\Omega R_i,5_i}$ and $\gamma_{\Omega R_1R_2,1}$ are
symmetric $32\!\times\!  32$ matrices. Hence there are 32 9-branes, 32 
$5_i$-branes ($i=1,2$) and 32 D-strings in total.

Before we give the explicit solution for the matrices $\gamma_{*,*}$, let us 
point out that the $\Omega^2$ eigenvalue of the oscillators in the $95_i$ 
and $5_i1$ strings is $-1$, for reasons discussed in \cite{gp}; 
while for the $91$ strings it is +1. 
This implies that since $\gamma_{\Omega,9}$ is symmetric 
$\gamma_{\Omega,1}$ must also be symmetric and 
$\gamma_{\Omega,5_i}$ are antisymmetric matrices. The rest of the
terms cancel with the choice of (anti-)symmetric traceless matrices
\begin{equation}
\begin{array}{| c | c c c c c c c |}
\hline
{}&\gamma_{\Omega}&\gamma_{\Omega R_1}& \gamma_{\Omega R_2}&
\gamma_{\Omega R_1R_2}&\gamma_{R_1}&\gamma_{R_2}&
\gamma_{R_1R_2}\\
\hline
9&\mbox{\bf 1}&M&N&-D&M&N&-D\\
5_1&M&D&\mbox{\bf 1}&N&N&M&-D\\
5_2&M&\mbox{\bf 1}&D&N&M&N&-D\\
1&\mbox{\bf 1}&M&N&-D&M&N&-D\\
\hline
\end{array}
\end{equation}
where, 
\begin{equation}
M=\left(\begin{array}{c c}
                      0&1\\
                     -1&0
            \end{array}\right),\quad
N=\left(\begin{array}{c c}
                      i\sigma_2&0\\
                     0&i\sigma_2
            \end{array}\right),\quad
D=-MN=\left(\begin{array}{c c}
                      0&-i\sigma_2\\
                     i\sigma_2&0
            \end{array}\right),
\end{equation}
$\sigma_2$ being the Pauli matrix. The above representation is schematic
in the sense that the size of the blocks depend on the number of branes
involved. 

\subsection{Solution of Chan-Paton Matrices and Anomaly Cancellation}

\noindent 99 {\bf Sector}: The `gauge group' on the 9-branes 
is $U(8)^2$. There are four antichiral fermions in the adjoint, and
another four in the $(\mbox{\bf 28}+\overline{\mbox{\bf 28}},\mbox{\bf 1})+
(\mbox{\bf 1},\mbox{\bf 28}+\overline{\mbox{\bf 28}})$, or equivalently in
$2(\mbox{\bf 28},\mbox{\bf 1})+2(\mbox{\bf 1},\mbox{\bf 28})$.
Eight antichiral bosons and two chiral multiplets of $(0,4)$ 
supersymmetry are in the $(\mbox{\bf 8},\mbox{\bf 8})+
(\overline{\mbox{\bf 8}},\overline{\mbox{\bf 8}})$. 

\medskip
\noindent $5_i5_i$ {\bf Sector}: In the first case ($\mbox{A}_i$), let there
be $4A_i$, ($A_i\le 8$), $5_i$-branes at a fixed plane\footnote{Due to 
the block structure
of the $\gamma$ matrices the branes move only as four-packs.}.  Then 
the `gauge group' on 
the $5_i$-branes is $U(A_i)^2$. There are four antichiral 
fermions in the adjoint, and another four in the 
$({1\over 2}\mbox{\bf A}_i(\mbox{\bf A}_i-\mbox{\bf 1})+\overline{{1\over 2}
\mbox{\bf A}_i(\mbox{\bf A}_i-\mbox{\bf 1})}, \mbox{\bf 1})+(\mbox{\bf 1},
{1\over 2}\mbox{\bf A}_i(\mbox{\bf A}_i-\mbox{\bf 1})+
\overline{{1\over 2}\mbox{\bf A}_i(\mbox{\bf A}_i-\mbox{\bf 1})})$.
Eight antichiral bosons and two chiral multiplets of (0,4) 
supersymmetry are in the $(\mbox{\bf A}_i,\mbox{\bf A}_i)+
(\overline{\mbox{\bf A}_i},\overline{\mbox{\bf A}_i})$. 

In the second case ($\mbox{B}_i$), when there are $4B_i$, (not counting
their images, which are at a different point of $T^4$), $5_i$-branes off the 
$R_i$ fixed planes the `gauge group' on the worldvolume is $U(2B_i)$, 
($B_i\le 4$).
Now there are four antichiral fermions, four antichiral bosons and one 
chiral multiplet in the adjoint. Four antichiral boson and one chiral 
multiplet are in the $\mbox{\bf B}_i(\mbox{\bf 2B}_i+\mbox{\bf 1})+ 
\overline{\mbox{\bf B}_i(\mbox{\bf 2B}_i+\mbox{\bf 1})}$; and four antichiral
fermions are in the $\mbox{\bf B}_i(\mbox{\bf 2B}_i-\mbox{\bf 1})+ 
\overline{\mbox{\bf B}_i(\mbox{\bf 2B}_i-\mbox{\bf 1})}$.

\medskip
\noindent 11 {\bf Sector}: In case (a), let there be $4a$, ($a\le 8$), 
D-strings at a fixed
for of ${\cal R}=R_1R_2$. The spectrum is same as in case A$_i$ above,
with the number $A_i$ replaced by $a$.

In case (b), (analogously (c)), with a bunch of $4b$, ($b\le 4$), 
D-strings, the `gauge
group' is $U(2b)$, and the spectrum is same as in case B$_i$ with $b$
replacing $B_i$ everywhere. 

When $d$ four-packs of D-strings, ($d\le 2$), are away from 
any fixed point, case (d),
the `gauge group' on their worldsheet is $SO(4d)$. All eight antichiral   
fermions are in the adjoint, eight antichiral bosons and two chiral 
multiplets are in the symmetric representation 
${\bf 2d}(\mbox{\bf 4d}+\mbox{\bf 1})$. 

\medskip
\noindent $95_i$ {\bf Sector}: There are four antichiral bosons, four 
antichiral fermions and one chiral multiplet in the 
$(\mbox{\bf 8},\mbox{\bf 1},\mbox{\bf A}_i,\mbox{\bf 1}) +
(\mbox{\bf 1},\mbox{\bf 8},\mbox{\bf 1},\mbox{\bf A}_i)$
representation of $U(8)^2\!\times\! U(A_i)^2$. Further one finds four 
antichiral bosons, four antichiral fermions and one chiral multiplet in the
$(\mbox{\bf 8},\mbox{\bf 1},\mbox{\bf 2B}_i) +
(\mbox{\bf 1},\mbox{\bf 8},\mbox{\bf 2B}_i)$
of $U(8)^2\!\times\! U(2B_i)$.

\medskip
\noindent 91 {\bf Sector}: There is only one antichiral fermion in this sector. 
It is in (a) 
$2(\mbox{\bf 8},\mbox{\bf 1},\mbox{\bf a},\mbox{\bf 1}) +
2(\mbox{\bf 1},\mbox{\bf 8},\mbox{\bf 1},\mbox{\bf a})$
of $U(8)^2\times U(a)^2$; (b)
$2(\mbox{\bf 8},\mbox{\bf 1},\mbox{\bf 2b}) +
2(\mbox{\bf 1},\mbox{\bf 8},\mbox{\bf 2b})$
of $U(8)^2\!\times\! U(2b)$; (c) similar to case (b); and (d)
$2(\mbox{\bf 8},\mbox{\bf 1},\mbox{\bf 4d}) +
2(\mbox{\bf 1},\mbox{\bf 8},\mbox{\bf 4d})$
of $U(8)^2\!\times\! SO(4d)$.

\medskip
\noindent $5_15_2$  {\bf Sector}: Once again there is only an antichiral
fermion. In case (A$_1$A$_2$) it is in 
$2(\mbox{\bf A}_1,\mbox{\bf 1},\mbox{\bf A}_2,\mbox{\bf 1}) +
2(\mbox{\bf 1},\mbox{\bf A}_1,\mbox{\bf 1},\mbox{\bf A}_2)$
of $U(A_1)^2\!\times\! U(A_2)^2$; case (A$_1$B$_2$) it is in 
$2(\mbox{\bf A}_1,\mbox{\bf 1},\mbox{\bf 2B}_2) +
2(\mbox{\bf 1},\mbox{\bf A}_1,\mbox{\bf 2B}_2)$
of $U(A_1)^2\!\times\! U(2B_2)$. Case (B$_1$A$_2$) is similar to  
(A$_1$B$_2$), and in case (B$_1$B$_2$) we have the antichiral fermion in
the $4(\mbox{\bf 2B}_1,\mbox{\bf 2B}_2)$
$SO(2B_1)\!\times\! SO(2B_2)$.

\medskip
\noindent $5_i1$ {\bf Sector}: In all configurations that lead to 
massless excitations, there are four antichiral bosons, four antichiral 
fermions and one chiral multiplet. They come in the
the following representations
\begin{list}{}{\itemindent 24pt}
\item[(A$_1$a)] 2({\bf A}$_1$,{\bf 1},{\bf a},{\bf 1})+2({\bf 1},{\bf A}$_1$,
{\bf 1},{\bf a}) of $U(A_1)^2\!\times\! U(a)^2$,
\item[(A$_1$b)] 2({\bf A}$_1$,{\bf 1},{\bf 2b})+2({\bf 1},{\bf A}$_1$,{\bf 2b}) of
$U(A_1)^2\!\times\! U(2b)$,
\item[(B$_1$a)] 2({\bf 2B}$_1$,{\bf a},{\bf 1})+2({\bf 2B}$_1$,{\bf 1},{\bf a}) of
$U(2B_1)\!\times\! U(a)^2$,
\item[(B$_1$c)] 4({\bf 2B}$_1$,{\bf 2c}) of $U(2B_1)\!\times\! U(2c)^2$,
\item[(B$_1$d)] 4({\bf 2B}$_1$,{\bf 4d}) of $U(2B_1)\!\times\! SO(4d)^2$;
\end{list}
and similarly for $5_21$ strings.
This completes the enumeration of the open string spectrum.

\medskip
If a four-pack of $5_i$-brane moves away from a $R_i$ fixed plane, its
image also moves. When a four-pack of D-strings moves 
away from an $R_i$ fixed plane, while remaining on the $R_{j\ne i}$ fixed
plane, it has one image, and away from all fixed planes there are three
images. This leads to the constraints
\begin{equation}
4\left( A_1 + 2B_1\right) = 4\left( A_2 + 2B_2\right) = 4\left( a + 2b+2c+4d\right)
=32,
\end{equation}
since there are 32 branes of each type in all. 

Once again, the numbers of chiral and antichiral bosons are equal, so
we need only to count the difference in the fermions in order to check for
the anomaly. The contributions from the various sectors are 
\begin{equation}
\begin{array}{| c || c | c  | c | c | c | c |}
\hline
\mbox{Sector}&\mbox{closed}&99&5_i5_i&11&91&5_15_2\\
\hline
f_+-f_-&-160&-64&-8A_i-16B_i&-8a-16b&
32a+64b&4\left(A_1+2B_1\right)\\
 & & & & -16c-32d &+64c+128d &\left(A_2+2B_2\right) \\
\hline
\end{array}
\end{equation}
The sectors not mentioned above do not have a net contribution to the
anomaly. The interested reader can easily check that the anomaly 
vanishing condition (\ref{ganomaly}) is satisfied for the $(0,4)$ 
supersymmetric $T^8/\zed_2^2$ orientifold. 

\section{The $T^8/\zed_2^3$ Orientifold}\label{zee2cub}

\subsection{Preliminaries}

The $\zed_2$ actions are generated by three reflections $R_1$, $R_2$ 
and $R_3$ which act on the coordinates of the torus $T^8$ as
\begin{equation}
\begin{array}{c c c c}
R_1:&\left(x^{2,3},x^{4,5},x^{6,7},x^{8,9}\right)&\to&
\left(x^{2,3},x^{4,5},-x^{6,7},-x^{8,9}\right)\\
R_2:&\left(x^{2,3},x^{4,5},x^{6,7},x^{8,9}\right)&\to&
\left(-x^{2,3},-x^{4,5},x^{6,7},x^{8,9}\right)\\
R_3:&\left(x^{2,3},x^{4,5},x^{6,7},x^{8,9}\right)&\to&
\left(-x^{2,3},x^{4,5},-x^{6,7},x^{8,9}\right)\\
\end{array}\label{reflex3}
\end{equation}
The torus $T^8$ identified by these reflections can be blown up to a
smooth fourfold of $SU(4)$ holonomy, which preserves one eighth of
the original supresymmetry. However, we also quotient by the worldsheet 
parity $\Omega$, which kills another half yielding a $(0,2)$ 
supersymmetric model in two dimensions.    

The group $\zed_2^3$ has altogether eight elements, six of which leave
16 O-5-planes invariant each while ${\cal R}=R_1R_2$ leads to 256 
O-1-planes as before. Denoting with a `$\cdot $' a direction in which 
a brane/plane is localized and with a `---' one in  which it is 
extended\footnote{All branes are extended along the non compact 
$x^{0,1}$ directions.}, we collect the branes and planes involved in
Table \ref{zed3branes}.
\begin{table}[h]
\begin{center}
\begin{tabular}{| c || c | c | c | c || c |}
\hline
Reflection&$x^{2,3}$&$x^{4,5}$&$x^{6,7}$&$x^{8,9}$&Resulting Branes\\
\hline
\hline
1&---&---&---&---&9\\
$R_1$&---&---&$\cdot$&$\cdot$&$5_1$\\
$R_2$&$\cdot$&$\cdot$&---&---&$5_2$\\
$R_3$&$\cdot$&---&$\cdot$&---&$5_3$\\
$R_1R_3\equiv R_4$&$\cdot$&---&---&$\cdot$&$5_4$\\
$R_2R_3\equiv R_5$&---&$\cdot$&$\cdot$&---&$5_5$\\
$R_1R_2R_3\equiv R_6$&---&$\cdot$&---&$\cdot$&$5_6$\\
$R_1R_2\equiv {\cal R}$&$\cdot$&$\cdot$&$\cdot$&$\cdot$&1\\
\hline
\end{tabular}
\end{center}
\caption{Branes and Planes of the $T^8/\zed_2^3$ Orientifold}
\label{zed3branes}
\end{table} 
Notice that the $5_1$-branes are orthogonal to the $5_2$-branes in
the sense that their worldvolumes extend along mutually orthogonal
four tori in $T^8$. Likewise $5_3$ and $5_6$, and $5_4$ and $5_5$ branes 
are mutually orthogonal pairs. 

To specify the reflections $R_1$, $R_2$ and $R_3$ on the Ramond sector 
states, we define them as
\begin{eqnarray}
R_1 & = & e^{i\pi\left( J_{67} - J_{89}\right)} \nonumber\\
R_2 & = & e^{i\pi\left( J_{45} - J_{23}\right)} \\
R_3 & = & e^{i\pi \left( J_{23} - J_{67}\right)} \nonumber
\end{eqnarray}
The supercharge $Q$ given by (\ref{superch}) is invariant under 
$\zed_2^3$.

\subsection{Massless Spectrum: Closed String Sector}: 

We begin with the right-movers from the untwisted sector of the closed
string. The worldsheet left movers together with their $R_1$, 
$R_2$ and $R_3$ eigenvalues are 
\begin{equation}
\begin{array}{r c l l c c c c}
\mbox{\underline{sector}}&{}&\mbox{\underline{state}}&{}&{}&
\mbox{\underline{$R_1$}}&\mbox{\underline{$R_2$}}
&\mbox{\underline{$R_3$}}\\
\mbox{NS}&{}&\psi^i_{-{1\over 2}}|0\rangle,&i=2,3&{}&+&-&-\\
{}&{}&\psi^i_{-{1\over 2}}|0\rangle,&i=4,5&{}&+&-&+\\
{}&{}&\psi^i_{-{1\over 2}}|0\rangle,&i=6,7&{}&-&+&-\\
{}&{}&\psi^i_{-{1\over 2}}|0\rangle,&i=8,9&{}&-&+&+\\
\mbox{R}&{}&|+,2s_1,\cdots,2s_4\rangle,&s_1=s_2=s_3=s_4&{}&+&+&+\\
{}&{}&|+,2s_1,\cdots,2s_4\rangle,&s_1=s_2=-s_3=-s_4&{}&+&+&-\\
{}&{}&|+,2s_1,\cdots,2s_4\rangle,&s_1=-s_2=s_3=-s_4&{}&-&-&+\\
{}&{}&|+,2s_1,\cdots,2s_4\rangle,&s_1=-s_2=-s_3=s_4&{}&-&-&-\\
\end{array}
\label{reigen3}
\end{equation}
Similarly for the worldsheet right-movers. The invariant combinations 
from the NS-NS and RR lead to $12+4=16$ antichiral bosons. 
$\Omega$ projects out antichiral fermions.
 
The left-movers obtained with the light-cone condition $p_+=0$, 
have the same eigenvalues in the NS sector as their right-moving
counterparts. The R sector states and their eigenvalues are
\begin{equation}
\begin{array}{r c l l c c c c}
\mbox{\underline{sector}}&{}&\mbox{\underline{state}}&{}&{}&
\mbox{\underline{$R_1$}}&\mbox{\underline{$R_2$}}&
\mbox{\underline{$R_3$}}\\
\mbox{R}&{}&|-,2s_1,\cdots,2s_4\rangle,&s_1=s_2=s_3=-s_4&{}&-&+&+\\
{}&{}&|-,2s_1,\cdots,2s_4\rangle,&s_1=s_2=-s_3=s_4&{}&-&+&-\\
{}&{}&|-,2s_1,\cdots,2s_4\rangle,&s_1=-s_2=s_3=s_4&{}&+&-&+\\
{}&{}&|-,2s_1,\cdots,2s_4\rangle,&-s_1=s_2=s_3=s_4&{}&+&-&-\\
\end{array}
\label{leigen3}
\end{equation}
Once again, we have 16 chiral bosons, but in addition there are also 16
chiral fermions. These can be combined into eight chiral multiplets of
$(0,2)$ supersymmetry. 

In the twisted sector massless modes come from the elements $R_1$,
$R_2$, $R_3$, $R_4=R_1R_3$, $R_5=R_2R_3$ and $R_6=R_1R_2R_3$ 
while ${\cal R}
= R_1R_2$ gives only massive excitations as in the previous models.
All the $R_i$-twisted sectors are similar, 
therefore we will only give the details 
for  the $R_1$ twisted sector. The right-movers and their $R_i$ 
eigenvalues are
\begin{equation}
\begin{array}{r c l l c c c c}
\mbox{\underline{sector}}&{}&\mbox{\underline{state}}&{}&{}&
\mbox{\underline{$R_1$}}&\mbox{\underline{$R_2$}}&
\mbox{\underline{$R_3$}}\\
\mbox{NS}&{}&|2s_3,2s_4\rangle,&s_3=s_4&{}&+&+&-2is_3\\
\mbox{R}&{}&|+,2s_1,2s_2\rangle,&s_1=-s_2&{}&+&-&+2is_1,\\
\end{array}
\label{treigen3}
\end{equation}
while the left-movers
\begin{equation}
\begin{array}{r c l l c c c c}
\mbox{\underline{sector}}&{}&\mbox{\underline{state}}&{}&{}&
\mbox{\underline{$R_1$}}&\mbox{\underline{$R_2$}}&
\mbox{\underline{$R_3$}} \\
\mbox{NS}&{}&|2s_3,2s_4\rangle,&s_3=s_4&{}&+&+&-2is_3\\
\mbox{R}&{}&|-,2s_1,2s_2\rangle,&s_1=s_2&{}&+&+&+2is_1.\\
\end{array}
\label{tleigen3}
\end{equation}
The invariant combinations are two antichiral bosons, two chiral
bosons and two chiral fermions from each fixed plane of $R_1$. The chiral
fields combine into one chiral multiplet per fixed plane.

To summarize, the closed string sector gives 208 antichiral bosons and
104 chiral multiplets of $(0,2)$ supersymmetry. 

\subsection{Open String Sector: Massless Spectrum and Configuration}

The discussion of the massless spectrum of the open string sector is
a straightforward extension of the ideas explained in Sections \ref{zed1open}
and \ref{zed2open}. Therefore we will not repeat it here. Instead we will
enumerate the various possible configurations of branes.

Let us begin with the 9-branes. Open strings with both ends on 9-branes
are subject to projections by $\Omega$ and all reflections $R_1, R_2$ and
$R_3$. If an open string has one end on the 9-brane and the other on a
5-brane (say the $5_1$-brane), then the position of the 5-brane determines
the relevant projections. (Strings with ends on different branes are not
constrained by $\Omega$.) Various possible configurations of the $95_1$
strings are given in Table \ref{ninefiveone}. 

\begin{table}[h]
$$
\begin{array}{| c || c | c || c || c |}
\hline 
\mbox{Case}&\multicolumn{2}{ c ||}{\mbox{At fixed point of}}
&\mbox{Projections}&\mbox{Located at fixed point along}\\
\cline{2-3}
{}&R_1R_3&R_3&{}&{}\\
\hline
\hline
\mbox{A}_1&\mbox{yes}&\mbox{yes}&R_1,\; R_2,\; R_3&x^{6,7},x^{8,9}\\
\mbox{B}_1&\mbox{yes}&\mbox{no}&R_2,\; R_1R_3&x^{8,9}\\
\mbox{C}_1&\mbox{no}&\mbox{yes}&R_2,\; R_3&x^{6,7}\\
\mbox{D}_1&\mbox{no}&\mbox{no}&R_2&\mbox{---}\\
\hline
\end{array}
$$
\caption{Configurations of $95_1$ strings} 
\label{ninefiveone}
\end{table}

All possible configurations of $5_15_1$ strings can be read from 
Table \ref{ninefiveone}. These are also restricted by the $\Omega$
condition. All other $95_i$ strings and $5_i5_i$, ($i=1,\cdots, 6$) strings 
are analogous. For $5_i5_j$, $i\ne j$, there are qualitatively two different 
types: the mutually orthogonal ones $5_15_2$, $5_35_6$ and $5_45_5$,
and the remaining ones which overlap along a $T^2$ subspace of $T^8$. 
The latter types do not always give rise to massless excitations. As an 
example, we display the possible configurations of $5_15_3$ strings 
and the projections to be imposed on them in
Table \ref{fiveonefivethree}.

\begin{table}[h]
\begin{center}
\begin{tabular}{| c || c | c | c | c |}
\hline
$5_35_1$&$\mbox{A}_1$&$\mbox{B}_1$&$\mbox{C}_1$&$\mbox{D}_1$\\
\hline\hline
$\mbox{A}_3$&$R_1,\; R_2,\; R_3$&(massive)&$R_2,\; R_3$&(massive)\\
\hline
$\mbox{B}_3$&$R_1,\; R_2R_3$&(massive)&$R_2R_3$&(massive)\\
\hline
$\mbox{C}_3$&(massive)&$R_2,\; R_1R_3$&(massive)&$R_2$\\
\hline
$\mbox{D}_3$&(massive)&$R_1R_2R_3$&(massive)&none\\
\hline
\end{tabular}
\end{center}
\caption{Configurations of $5_15_3$ Strings}
\label{fiveonefivethree}
\end{table}

For the 1-branes there are seven different possibilities depending on its
location at fixed planes of one, two or three reflections. We will see that in
this model branes are allowed to move in eight-packs only, and that
they are altogether 32 in number. This prevents the D-strings to 
move away from all fixed planes since this would require that their seven
eight-packs of images move as well. The configurations of 91 strings are
summarized in Table \ref{nineone}, from which it is possible to determine
all possible 11 configurations. 

\begin{table}[h]
$$
\begin{array}{| c || c | c | c || c || c |}
\hline 
\mbox{Case}&\multicolumn{3}{ c ||}{\mbox{At fixed point of}}&\mbox{Projections}
&\mbox{Location at fixed point along}\\
\cline{2-4}
{}&R_1R_3&R_2&R_3&{}&{}\\ 
\hline
\hline
\mbox{a}&\mbox{yes}&\mbox{yes}&\mbox{yes}&R_1,\; R_2,\; R_3
&x^{2,3},x^{4,5},x^{6,7},x^{8,9}\\
\mbox{b}&\mbox{yes}&\mbox{yes}&\mbox{no}&R_1R_3,\; R_2&x^{2,3},
x^{4,5},x^{8,9}\\
\mbox{c}&\mbox{yes}&\mbox{no}&\mbox{yes}&R_1R_3,\; R_3
&x^{2,3},x^{6,7},x^{8,9}\\
\mbox{d}&\mbox{no}&\mbox{yes}&\mbox{yes}&R_2,\; R_3
&x^{2,3},x^{4,5},x^{6,7}\\
\mbox{e}&\mbox{yes}&\mbox{no}&\mbox{no}&R_1R_3&x^{2,3},x^{8,9}\\
\mbox{f}&\mbox{no}&\mbox{yes}&\mbox{no}&R_2&x^{2,3},x^{4,5}\\
\mbox{g}&\mbox{no}&\mbox{no}&\mbox{yes}&R_3&x^{2,3},x^{6,7}\\
\hline
\end{array}
$$
\caption{Configurations of 91 strings} 
\label{nineone}
\end{table}

Finally, the various configurations of the $5_i1$ strings can be obtained
by combining the $95_i$ sector with the $5_i1$ sector. One needs to keep
in mind that in many configurations the $5_i$-brane and the D-string are
separated in space leading only to massive excitations.

\subsection{Tadpole Cancellation}

We will not display the tadpole divergence from the various diagrams 
explicitly as they are analogus to the previous models. The expressions
are however longer because now there are six different types of 5-branes. 
Demanding that the total divergence adds up to zero requires that there
are 32 branes of each type, and that $\gamma_{\Omega,9}$, 
$\gamma_{\Omega R_i,5_i}$, ($i=1,\cdots, 6$), and 
$\gamma_{\Omega{\cal R},1}$ are symmetric matrices. Additionally, the
matrices $\gamma_{R,*}$ for any reflection $R$ must be traceless in 
all sectors.  

To find the solution for $\gamma_{*,*}$, we need to take into account
the eigenvalues given in Table \ref{phase} in Appendix \ref{phases3}. 
The explicit solutions are given in Tables \ref{soln1} and \ref{soln2}. 
The matrices $s_i$ and $a_i$ appearing in the tables are defined in 
Appendix \ref{matrices3}.

\begin{table}[h]
$$
\begin{array} {| c || c | c | c | c | c | c | c | c|}
\hline
 & \gamma_\Omega &  \gamma_{\Omega R_1} &  \gamma_{\Omega R_2} &
 \gamma_{\Omega R_3} &  \gamma_{\Omega R_1R_2R_3 } & 
 \gamma_{\Omega R_2 R_3} &  \gamma_{\Omega R_1 R_3} &
\gamma_{\Omega R_1R_2}\\
\hline \hline
9 & 1 & a_1 & a_2 & a_5 &- a_6 & a_7 & a_8 &- s_3 \\
\hline
5_1 & a_3 &- s_1 &- s_2 & a_5 &- a_6 & a_7 & a_8 &- a_4 \\
\hline
5_2 & a_3 &- s_1 &- s_2 & a_5 &- a_6 & a_7 & a_8 &- a_4\\
\hline
5_3 & a_3 & a_5 & a_6 &- s_1 & -s_2 & -a_7 & -a_8 &a_4\\
\hline
5_4 & a_3 & a_5 & a_6 & a_7 & a_8 & s_1 & s_2 &a_4\\
\hline
5_5 & a_3 & a_5 & a_6 & a_7 & a_8 & s_1 & s_2 & a_4\\
\hline
5_6 & a_3 & a_5 & a_6 &- s_1 &- s_2 &- a_7 &- a_8 &a_4\\
\hline
1 & 1 & a_1 & a_2 & a_5 &- a_6 & a_7 & a_8 &- s_3 \\
\hline
\end{array}
$$
\caption{Explicit solution to projection matrices I}
\label{soln1}
\end{table}
\begin{table}[h]
$$
\begin{array}  {| c || c | c | c | c | c | c | c| }
\hline
 &  \gamma_{ R_1} &  \gamma_{ R_2} &
 \gamma_{ R_3} &  \gamma_{ R_1R_2R_3 } & 
 \gamma_{ R_2 R_3} &  \gamma_{ R_1 R_3} &
\gamma_{R_1 R_2}\\
\hline \hline
9     & a_1 & a_2 & a_5 &- a_6 & a_7 & a_8 &-s_3\\ \hline
5_1 & a_1 & a_2 & s_4 &- s_5 & s_6 & s_7 &-s_3 \\ \hline
5_2 & a_1 & a_2 & s_4 & -s_5 & s_6 & s_7 &-s_3\\ \hline
5_3 & s_4 & s_5 & a_1 & a_2 & -s_6 & -s_7 &s_3\\ \hline
5_4 & s_4 & s_5 & s_6 & s_7 &- a_1 &- a_2 &s_3\\ \hline
5_5 & s_4 & s_5 & s_6 & s_7 &- a_1 &- a_2 &s_3\\ \hline
5_6 & s_4 & s_5 & a_1 & a_2 &- s_6 & -s_7 &s_3\\ \hline
1     & a_1 & a_2 & a_5 &- a_6 & a_7 & a_8 &-s_3\\ \hline 
\end{array} 
$$
\caption{Explicit solution to projection matrices II.}
\label{soln2}
\end{table}

\subsection{Solution of Chan-Paton Matrices and Anomaly Cancellation}

The representation matrices $\gamma_{*,*}$ have an $8\!\times\! 8$ 
block structure. This implies that the branes can move only as
eight-packs. In the following we first catalog the massless excitations 
coming from various sectors and then check for the anomaly 
condition. 

\medskip
\noindent 99 {\bf Sector}: The `gauge group' is $Sp(8)\!\times\! Sp(8)$. 
Two antichiral
fermions are in the adjoint ({\bf 36},{\bf 1})+({\bf 1},{\bf 36}), six more
antichiral fermions are in the (reducible) antisymmetric rank two 
({\bf 28},{\bf 1})+({\bf 1},{\bf 28}) representation. Eight antichiral bosons
and four chiral multiplets are in the ({\bf 8},{\bf 8}) representation.  

\medskip
\noindent $5_i5_i$ {\bf Sector}: In case (A$_i$) the `gauge group' on the
$5_i$-brane is $Sp(2A_i)^2$, ($A_i\le 4$). 
Two antichiral fermions are in the adjoint
({\bf A}$_i$({\bf 2A}$_i$+{\bf 1}),{\bf 1})+({\bf 1},{\bf A}$_i$({\bf 2A}$_i$+{\bf 1})).
Six antichiral fermions are in the 
({\bf A}$_i$({\bf 2A}$_i$$-${\bf 1}),{\bf 1})+
({\bf 1},{\bf A}$_i$({\bf 2A}$_i$$-${\bf 1})),
eight antichiral bosons and four chiral multiplets are in the 
({\bf 2A}$_i$,{\bf 2A}$_i$).

In case (B$_i$), the `gauge group' is $Sp(4B_i)$, ($B_i\le 2$), 
two antichiral
fermions, six antichiral bosons and three chiral multiplets are all in 
the adjoint; six antichiral fermions and one chiral multiplet
are in the antisymmetric 
representation {\bf 2B}$_i$({\bf 4B}$_i$$-${\bf 1}). Case (C$_i$) is similar. 

In case (D$_i$), when the $5_i$-branes are off all fixed planes, the `gauge
group' is $U(4D_i)$, ($D_i\le 1$). 
There are four antichiral fermions in the adjoint
and four antichiral fermions in 
the {\bf 2D}$_i$({\bf 4D}$_i$$-${\bf 1}) + 
$\overline{\mbox{{\bf 2D}$_i$({\bf 4D}$_i$$-${\bf 1})}}$
representation. Further there are
four antichiral bosons and two chiral multiplets in the adjoint,
four antichiral bosons and two chiral multiplets in the 
{\bf 2D}$_i$({\bf 4D}$_i$+{\bf 1}) + 
$\overline{\mbox{{\bf 2D}$_i$({\bf 4D}$_i$+{\bf 1})}}$
representation.

\medskip
\noindent 11 {\bf Sector}: There are several cases to consider here. In case
(a) with $8a$ D-strings, ($a\le 4$), at the 
fixed point of ${\cal R}$, the spectrum is
similar to case (A$_i$) of $5_i5_i$ strings with $A_i$ replaced by $a$. 
Case (b) again is obtained from the case (B$_i$) with $B_i$ substituted
by $b$. The cases (c) and (d) are similar to (b). To get the spectrum in 
cases (e), (f) and (g) replace $D_i$ in case (D$_i$) above by $e$, 
(respectively by $f$ and $g$).

\medskip
\noindent $95_i$ {\bf Sector}:  Here two antichiral bosons, two antichiral
fermions and one chiral multiplet all transform in the same way.
In different configurations these are ({\bf 8},{\bf 1},{\bf 2A}$_i$,{\bf 1})
+ ({\bf 1},{\bf 8},{\bf 1},{\bf 2A}$_i$) of $Sp(8)^2\!\times Sp(2A_i)^2$;
({\bf 8},{\bf 1},{\bf 4B}$_i$)+({\bf 1},{\bf 8},{\bf 4B}$_i$) 
of $Sp(8)^2\!\times\! Sp(4B_i)$; and 
2({\bf 8},{\bf 1},{\bf 4D}$_i$)+2({\bf 1},{\bf 8},{\bf 4D}$_i$)
of $Sp(8)^2\!\times\! U(4D_i)$ respectively.

\medskip
\noindent 91 {\bf Sector}: The only massless excitation is an antichiral 
fermion. This is in the representation (a) ({\bf 8},{\bf 1},{\bf 2a},{\bf 1})
+ ({\bf 1},{\bf 8},{\bf 1},{\bf 2a}) of $Sp(8)^2\!\times Sp(2a)^2$; (b)
({\bf 8},{\bf 1},{\bf 4b})+({\bf 1},{\bf 8},{\bf 4b}) of $Sp(8)^2\!\times\! Sp(4b)$
(similarly cases (c) and (d)); and 2({\bf 8},{\bf 1},{\bf 4e})+2({\bf 1},{\bf 8},
{\bf 4e}) of $Sp(8)^2\!\times\! U(4e)$, (cases (f) and (g) are similar).

\medskip
\noindent $5_i5_j$ {\bf Sector}: As has been discussed earlier, there are
two qualitatively different kinds of $5_i5_j$ strings. The first kind stretch
between mutually orthogonal pairs of branes ($5 _15_2$, $5_35_6$ 
and $5_45_5$) and contribute only one antichiral fermion. 
This is in the
\begin{list}{}{\itemindent 24pt}
\item [(A$_1$A$_2$)] ({\bf 2A}$_1$,{\bf 1},{\bf 2A}$_2$,{\bf 1})+({\bf 1},
{\bf 2A}$_1$,{\bf 1},{\bf 2A}$_2$) of $Sp(2A_1)^2\!\times Sp(2A_2)^2$; 
\item[(A$_1$B$_2$)] ({\bf 2A}$_1$,{\bf 1},{\bf 4B}$_2$)+({\bf 1},{\bf 2A}$_1$,
{\bf 4B}$_2$) of $Sp(2A_1)^2\!\times Sp(4B_2)$;
\item[(A$_1$D$_2$)] 2({\bf 2A}$_1$,{\bf 1},{\bf 4D}$_2$)+2({\bf 1},
{\bf 2A}$_1$,
{\bf 4D}$_2$) of $Sp(2A_1)^2\!\times\! U(4D_2)$;
\item[(B$_1$B$_2$)] 2({\bf 4B}$_1$,{\bf 4B}$_2$) of $Sp(4B_1)\!\times
\! Sp(4B_2)$;
\item[(B$_1$D$_2$)] 4({\bf 4B}$_1$,{\bf 4D}$_2$) of $Sp(4B_1)\!\times\!
U(4D_2)$;
\item[(D$_1$D$_2$)] 4({\bf 4D}$_1$,{\bf 4D}$_2$) of $U(4D_1)\!\times\! 
U(4D_2)$;
\end{list}
and so on. Only representative cases are displayed here. 
 
The second kind of $5_i5_j$ strings are stretched between nonothrogonal
branes that overlap along a two-torus $T^2$ in $T^8$. Consider for
example the $5_15_3$ strings --- not all 
configurations contribute to the massless spectrum (see Table 
\ref{fiveonefivethree}). When there are massless modes they are two
antichiral bosons, two antichiral fermions and one chiral multiplet in
the 
\begin{list}{}{\itemindent 24pt}
\item[(A$_1$A$_3$)] ({\bf 2A}$_1$,{\bf 1},{\bf 2A}$_3$,{\bf 1})+({\bf 1},
{\bf 2A}$_1$,{\bf 1},{\bf 2A}$_3$) of $Sp(2A_1)^2\!\times Sp(2A_3)^2$;
\item[(A$_1$B$_3$)] ({\bf 2A}$_1$,{\bf 1},{\bf 4B}$_3$)+({\bf 1},{\bf 2A}$_1$,
{\bf 4B}$_3$) of $Sp(2A_1)^2\!\times Sp(4B_3)$;
\item[(B$_1$C$_3$)] ({\bf 4B}$_1$,{\bf 4C}$_3$) of $Sp(4B_1)\!\times
\! Sp(4C_3)$;
\item[(B$_1$D$_3$)] 2({\bf 4B}$_1$,{\bf 4D}$_3$) of $Sp(4B_1)\!\times\!
U(4D_3)$;
\item[(D$_1$D$_3$)] 4({\bf 4D}$_1$,{\bf 4D}$_3$) of $U(4D_1)\!\times\! 
U(4D_3)$.
\end{list}
Once again we have discussed only representative classes --- the
rest being similar to one or another of the cases above. 

\medskip
\noindent $5_i1$ {\bf Sector}: As an example let us discuss the
$5_11$ sector. Once again different cases arise of which
many give rise to massive excitations only. In cases there are massless
modes they come as two antichiral bosons, two antichiral
fermions and one chiral multiplet in the
\begin{list}{}{\itemindent 24pt}
\item[(A$_1$a)] ({\bf 2A}$_1$,{\bf 1},{\bf 2a},{\bf 1})+({\bf 1},{\bf 2A}$_1$,
{\bf 1},{\bf 2a}) of $Sp(2A_1)^2\!\times Sp(2a)^2$;
\item[(A$_1$c)] ({\bf 2A}$_1$,{\bf 1},{\bf 4c})+({\bf 1},{\bf 2A}$_1$,{\bf 4c}) 
of $Sp(2A_1)^2\!\times\! Sp(4c)$, (similarly (B$_1$b) and (C$_1$d));
\item[(B$_1$e)] 2({\bf 4B}$_1$,{\bf 4e} of $Sp(4B_1)\!\times\! U(4e)$,
(similarly (C$_1$g) and (D$_1$f)).
\end{list}
In the cases not mentioned above, the lightest excitations are massive.

This closes the description of the open string sector. 

Altogether we have $32$ fivebranes of each type. Further when we
move an eight-pack of $5_i$-branes away from a fixed plane its image also
moves away leading to the constraint
\begin{equation}
8(A_i + 2B_i + 2C_i + 4D_i) = 32.
\label{fiveno}
\end{equation}
Simalarly for the D-strings we obtain a constraint on their numbers in
various positions
\begin{equation}
8(a + 2b + 2c +2d +4e +4f + 4g) = 32. 
\label{oneno}
\end{equation}
We recall here that no eight-pack of D-strings can escape all the fixed 
points since there are not enough of them to account for  the necessary
images.

Now we can compute the contributions to the anomaly due to the 
mismatch in the fermion numbers from various sectors. This is
displayed in Table \ref{mismatch}.
\begin{table}[h]
$$
\begin{array}{| c | c | c |}
\hline
\mbox{Sector} & f_+ - f_- & \mbox{Multiplicity}\\
\hline
\mbox{Closed}  & -208 & 1  \\
99 & -32 & 1\\
5_i5_i & -8A_i-16B_i-16C_i -32D_i & 6 \\
11 & -8a -16\left(b+c+d\right) -32\left(e+f+g\right) & 1 \\
91 & 32a + 64\left( b+c+d\right) + 128\left( e+f+g\right)  & 1\\
5_i\perp 5_j & 8\left(A_i + 2B_i +2C_i + 4D_i\right) & 3\\
  & \left(A_j + 2B_j + 2C_j + 4D_j\right)  & \\
\hline
\end{array} 
$$
\caption{Fermion mismatch in the $T^8/\zed_2^3$ orientifold}
\label{mismatch}
\end{table}
In sectors not mentioned in Table \ref{mismatch}, the number of 
chiral fermions matches with the number of antichiral fermions. 
The multiplicities are due to the six different kinds of fivebranes, 
and  $5_i\perp 5_j$ are pairs of fivebranes which are mutually orthogonal
in the compact directions. As mentioned earlier, there are three 
such pairs $5_1\perp 5_2$, $5_3\perp 5_6$, $5_4\perp 5_5$. 

Taking the difference in the fermion numbers from Table \ref{mismatch} and 
using Eqs.(\ref{fiveno}) and (\ref{oneno}), one can check the anomaly
vanishing condition (\ref{ganomaly}) is satisfied in the $(0,2)$ model.

\newpage
\section{Some Other Orientifolds of $T^8$}\label{others}

\subsection{The $T^8/\zed_2^4$ Orientifold}

The discrete group $\zed_2^4$ is generated by
the three reflections (\ref{reflex3}), and also in addition
\begin{equation}
\left(x^2,x^3,x^4,x^5,x^6,x^7,x^8,x^9\right)\to
\left(x^2,-x^3,x^4,-x^5,x^6,-x^7,x^8,-x^9\right)
\end{equation}
acting on the coordinates of $T^8$. This group is generated by 16 
elements, and the resulting space is an orbifold limit of an 
eight dimensional Joyce manifold of {\it Spin}(7) holonomy\cite{joyce}. 
There is one covariantly constant spinor in a Joyce manifold. Therefore type 
IIB string on it, or its orbifold limit $T^8/\zed_2^4$, leads to $(0,2)$ 
supersymmetry in two dimensions. 

If in addition, we quotient by the worldsheet parity $\Omega$ the 
resulting model has $(0,1)$ supersymmetry. There are also 9-branes,
1-branes and a host of 5-branes wrapping different $T^4$ 
subspaces of $T^8$. 
This model has the minimal supersymmetry and should be interesting
to study in detail. Unfortunately the construction is also rather tedious
and we will not carry out the details here. 

\subsection{T-dual of Type I String}

So far we have not discussed what is perhaps the simplest orientifold of
type IIB. This model is obtained from a quotient of type IIB on $T^8$ by
$\zed_2=\{1,\Omega{\cal R}\}$. The reflection ${\cal R}$ on the coordinates
of $T^8$ always occurs in combination with worldsheet parity reversal.
This model is however not chiral and hence not in the same league of models
described in the present paper. Indeed it has (8,8) supersymmetry and
leads to 32 D-strings, as can be checked by explicit computation. The
same conclusion is reached when one observes that this model is just
the T-dual of ten dimensional type I string compactified on $T^8$. 

\section{Conclusions}\label{conclu}

We have presented two dimensional models with chiral $(0,N)$, $N=2,4,8$,
supersymmetry from orientifolds of type IIB theory on an eight-torus. In all
cases there are D-branes in the background whose configurations we
describe. The massless excitations from the closed and open string sectors
are determined. This is used to show that the gravitational anomaly 
vanishes leading to consistent vacua. 

An interesting orientifold can be constructed from the $(T^4/\zed_2)\!\times
(T^4/\zed_2)$ model with (0,4) supersymmetry. If we choose the $T^4$'s
to be identical, we can quotient by an exchange symmetry of the two. We
expect the resulting model to have (0,3) supersymmetry, but whether it is
a consistent background can only be confirmed by an explicit calculation.
It may also be interesting to investigate the corresponding orbifold which is
a degenerate limit of the symmetric product of two K3 surfaces\cite{k3sq}.  

A particular limit of the models we study can be related to four dimensional
spacetime with cosmic strings. To this end, one chooses two radii of $T^8$
`large' as in \cite{bch}. However unlike the orbifold models of \cite{bch},
interesting new possibilities arise in case of the orientifolds. Firstly, the 
cosmic strings are dynamical since they are $D$-branes. Moreover 
(depending on the choice of large directions) there can be domain walls.
To see this take for example the $T^8/\zed_2^2$ orientifold of Section
\ref{zee2sq}, and choose `large' radii of compactification along $x^{5,6}$.
The fivebranes then behave as two dimensional extended objects in four
spacetime dimensions. 

To conclude, an interesting problem to look at is the nonperturbative limit 
of the models presented here. One approach is through 
F-theory\cite{vafaf} which 
is type IIB with varying dilaton-axion in a sevenbrane background. The 
general prescription to relate an orientifold to F-theory is described in 
Ref.\cite{sennonp}. In conventional F-theory vacua the self-dual 4-form 
potential is zero and hence there is no threebrane (see however \cite{svw}). 
A T-duality along two compact directions (say $x^{8,9}$) transforms the 
ninebranes and some fivebranes of our model into sevenbranes, but in 
addition the D-strings and the rest of the fivebranes become 
threebranes. Thus we must necessarily have a non-trivial background 
of the 4-form. 
Finally it would be interesting to see if the M-theory
orientifolds of Ref.\cite{chiral2d} with chiral supersymmetry in two
dimensions are related to the IIB orientifolds discussed here. 
 
\bigskip
\bigskip
\noindent {\bf Acknowledgement}: It is a pleasure to thank Sunil Mukhi and 
Stefan Theisen for valuable discussion. S.~F.\ is supported by GIF, the
German Israeli Foundation for Scientific Research, and D.~G.\ is 
supported by a Fellowship from the Alexander von Humboldt Foundation.
The work is also supported in part by TMR program ERBFMX-CT96-0045.

\newpage
\begin{appendix}

\section{Matrices for the $T^8/\zed_2^3$ Orientifold}\label{matrices3}

The matrices that we need to solve for the representation $\gamma_{*,*}$ 
of $\Omega$ and $R_i$ in various sectors are the following set given
below. Each is an $8n\!\times\! 8n$ matrix with $8n \leq 32$ being the 
number of branes moving together. 
$$
\begin{array}{c c c}
\!\! s_1 = \left( \begin{array} {c c c c}
0 &  0 &  0 & \! i\sigma_2\! \\
0 & 0 &\! i\sigma_2\! & 0 \\
0 &\!\!\! -i\sigma_2\! & 0 & 0\\
\!\!\! -i\sigma_2\! & 0 & 0 & 0 \end{array} \right)
&\!\! s_2 = \left( \begin{array} {c c c c}
0 &  0 &  0 &\! i\sigma_2\! \\
0 & 0 &\!\!\! - i\sigma_2\! & 0 \\
0 &\! i\sigma_2\! & 0 & 0\\
\!\!\! -i\sigma_2\! & 0 & 0 & 0 \end{array} \right)
&\!\! s_3 = \left( \begin{array} {c c c c}
1 &  0 &  0 & 0 \\
0 & -1 & 0 & 0 \\
0 & 0 & -1 & 0\\
0 & 0 & 0 & 1 \end{array} \right) \\
{}&{}&{}\\
\!\! s_4 = \left( \begin{array} {c c c c}
\sigma_1 &  0 &  0 & 0 \\
0 & \sigma_1 & 0 & 0 \\
0 & 0 & \sigma_1 & 0\\
0 & 0 & 0 & \sigma_1 \end{array} \right) 
&\!\! s_5 = \left( \begin{array} {c c c c}
\sigma_1 &  0 &  0 & 0 \\
0 &\!\! -\sigma_1 & 0 & 0 \\
0 & 0 &\!\! -\sigma_1 & 0\\
0 & 0 & 0 & \sigma_1 \end{array} \right) 
&\!\! s_6 = \left( \begin{array} {c c c c}
\sigma_3 &  0 &  0 & 0 \\
0 & \sigma_3 & 0 & 0 \\
0 & 0 & \sigma_3 & 0\\
0 & 0 & 0 & \sigma_3 \end{array} \right) \\
{}&{}&{}\\
\!\! s_7 = \left( \begin{array} {c c c c}
\sigma_3 &  0 &  0 & 0 \\
0 &\!\! -\sigma_3 & 0 & 0 \\
0 & 0 &\!\! -\sigma_3 & 0\\
0 & 0 & 0 & \sigma_3 \end{array} \right)
&\!\! a_1 = \left( \begin{array} {c c c c}
\!i\sigma_2\! &  0 &  0 & 0 \\
0 &\!\!\! - i\sigma_2\! & 0 & 0 \\
0 & 0 &\!\!\! -i\sigma_2\! & 0 \\
0 & 0 & 0 &\! i\sigma_2\! \end{array} \right)
&\!\! a_2 = \left( \begin{array} {c c c c}
\!i\sigma_2 &  0 &  0 & 0 \\
0 & i\sigma_2 & 0 & 0 \\
0 & 0 & i\sigma_2 & 0\\
0 & 0 & 0 & i\sigma_2 \end{array}\! \right) \\
{}&{}&{}\\
\!\! a_3 = \left( \begin{array} {c c c c}
0 &  0 &  0 & -1 \\
0 & 0 & 1 & 0 \\
0 & -1 & 0 & 0\\
1 & 0 & 0 & 0 \end{array} \right)
&\!\! a_4 = \left( \begin{array} {c c c c}
0 &  0 &  0 & -1 \\
0 & 0 &- 1 & 0 \\
0 & 1 & 0 & 0\\
1 & 0 & 0 & 0 \end{array} \right)
&\!\! a_5 = \left( \begin{array} {c c c c}
0 &  0 &  0 &\!\! -\sigma_1 \\
0 & 0 & \sigma_1 & 0 \\
0 &\!\! -\sigma_1 & 0 & 0\\
\sigma_1 & 0 & 0 & 0 \end{array} \right) \\
{}&{}&{}\\
\!\! a_6 = \left( \begin{array} {c c c c}
0 &  0 &   0 & \!\! -\sigma_1 \\
0 & 0 & \!\! -\sigma_1 & 0 \\
0 & \sigma_1 & 0 & 0\\
\sigma_1 & 0 & 0 & 0 \end{array} \right)
&\!\! a_7 = \left( \begin{array} {c c c c}
0 &  0 &  0 & \!\! -\sigma_3 \\
0 & 0 &\sigma_3 & 0 \\
0 & \!\! -\sigma_3 & 0 & 0\\
\sigma_3 & 0 & 0 & 0 \end{array} \right)
&\!\! a_8 = \left( \begin{array} {c c c c}
0 &  0 &  0 & \!\! -\sigma_3 \\
0 & 0 & \!\! -\sigma_3 & 0 \\
0 & \sigma_3 & 0 & 0\\
\sigma_3 & 0 & 0 & 0 \end{array} \right)
\end{array}
$$
In the above $\sigma_i$ are Pauli matrices, and each entry is to be
understood as a direct product with the $1_{n\times n}$ matrix. 

\newpage

\section{Phase Factors in the $T^8/\zed_2^3$ Orientifold}\label{phases3}

The oscillators of open strings with two ends on different branes have 
eigenvalue $\pm 1$ for $\Omega^2$ and squares of the reflections 
in $\zed_2^3$. These are summarized in the Table \ref{phase}. 

\begin{table}[h]
$$
\begin{array}{| c || c | c | c | c | c | c | c |c |}
\hline 
 & \Omega^2 & R_1 ^2 & R_2 ^2 & R_3 ^2 & \left( R_1R_2R_3\right)^2 &
\left( R_2R_3\right)^2 & \left( R_1R_3\right)^2 & \left( R_1 R_2\right) ^2 \\
\hline \hline
95_1/95_2 & - & + & + & - & - & - & - & + \\
\hline
95_3/95_6 & - & - & - & + & + & - & -& +\\
\hline
95_4/95_5 & - & - & - & - & - & + & + & +\\
\hline
91 & + & + & + & + & + & + & + & + \\
\hline
5_1 5_2/5_35_6/5_45_5 & + & + & + & + & + & + & +& +  \\
\hline
5_1 5_3/5_15_6 & + & - & - & - & - & + & +& +\\
\hline
5_15_4/5_15_5  & + & - & - & + & + & - & -& +\\
\hline
5_25_3/5_25_6 & + & - & - & - & - & + & +&+\\
\hline
5_25_4/5_25_5 & + & - & - & + & + & - & -&+\\
\hline
5_35_4/5_35_5 & + & + & + & - & - & - & -&+\\
\hline
5_45_6/5_55_6 & + & + & + & - & - & - & -&+\\
\hline
5_11/5_21 & - & + & + & - & - & - & -&+\\
\hline
5_31/5_61 & - & - & - & + & + & - & -&+\\
\hline
5_41/5_51 & - & - & - & - & - & + & +&+\\
\hline
\end{array}
$$
\caption{Phase factors of DN wave functions.} \label{phase}
\end{table}
These phases have to be taken into account in finding the representations
$\gamma_{*,*}$ on different sectors, as in \cite{gp}.

\end{appendix}

\end{document}